\documentclass[aps,twocolumn,showpacs,preprintnumbers,amsmath,amssymb,nofootinbib,superscriptaddress]{revtex4-1}

\usepackage{graphicx}
\usepackage{dcolumn}% Align table columns on decimal point
\usepackage{bm}% bold math%
\usepackage{amsmath}
\usepackage{hyperref}
\usepackage{color}
\def\mk{\mathbf{k}}

\def\ms{\mathbf{s}}
\def\BOmega{\boldsymbol{\Omega}}

\def\Bsigma{\boldsymbol{\sigma}}
\def\BSigma{\boldsymbol{\Sigma}}

\newcommand{\secref}[1]{Sec.~\ref{#1}}
\newcommand{\eqnref}[1]{Eq.~(\ref{#1})}% or {(\ref{#1})}

\begin{document}

\title{The semiclassical theory for spin dynamics in a disordered system}

\author{Tsung-Wei Chen}
\email{twchen@mail.nsysu.edu.tw}\affiliation{Department of
Physics, National Sun Yat-sen University, Kaohsiung 80424, Taiwan}

\author{Hsiu-Chuan Hsu}
\email{hcjhsu@nccu.edu.tw}\affiliation{Graduate Institute of Applied Physics, National Chengchi University, Taipei 11605, Taiwan}
\affiliation{Department of Computer Science, National Chengchi University, Taipei 11605, Taiwan}

\date{\today}

\begin{abstract}
We investigate the Drude model of spin dynamics in two-dimensional spin-orbit coupled systems. In the absence of an applied electric field, the spin aligns with the k-dependent effective magnetic field. The influence of disorder (the momentum relaxation time $\tau$) on the system is considered. In the presence of an electric field, the change in momentum causes a change in the effective magnetic field. The in-plane spin can precess around the successive change in the orientation of the effective magnetic field. We find that up to the linear order of the electric field, the spin orientation undergoes Larmor-like and non-Larmor like precession. Furthermore, we find that the non-Larmor motion over a very short time ($t\ll\tau$) exactly equals the result obtained from the Kubo formula. This means that the Kubo formula only captures the system's response over a very short evolution. The spin-Hall conductivity for Larmor and non-Larmor precession in the Rashba system is analytically calculated. We find that the intrinsic spin–Hall current is not a universal constant and correctly drops to zero when the Rashba spin-orbit coupling drops to zero. We also calculate the time-averaged Larmor and non-Larmor spin-Hall conductivities (SHCs) for the k-cubic Rashba system and compare them to experimental values. The time-averaged Larmor SHC vanishes and the non-Larmor SHC is given by $2.1(q/8\pi)$ which is very close to the experimental value $2.2(q/8\pi)$ by Wunderlich et al. [Phys. Rev. Lett. {\bf 94}, 047204 (2005)].

\end{abstract}
 \maketitle

\section{Introduction}
When applying an external electric field in a spin-orbit coupled system, lateral spin motion is generated and spin accumulates at the edges of the sample, which is known as the spin-Hall effect~\cite{Hirsch1999}. The spin-Hall effect has been extensively studied theoretically and through experiments over the past decades~\cite{Sinova2015}. The mechanism of the intrinsic spin–Hall effect stems from the Berry curvature~\cite{Mura2003, Chang2010} or the adiabatic evolution of spin around the effective magnetic field~\cite{Sinova2004}. A challenging issue is how to define the spin current because each component of spin is not a conserved quantity. In Ref.~\cite{Shi2006}, a generic definition of spin current (SZXN spin current) is proposed, and the spin current is composed of the conventional definition of spin current and the spin-torque dipole current. The conventional and spin-torque dipole currents were found to be related to the Berry phase in k-linear systems~\cite{Shen2004,Chen2014}, while no such relationship exists for spin-orbit coupled systems with higher order momentum. Finally, the spin-Hall conductivity obtained from the SZXN spin current in response to the electric field is a universal constant in the k-linear system, regardless of the strength of the spin-orbit coupling, such as in pure Rashba~~\cite{Rashba1984,Sinova2004} and Rashba–Dresselhaus systems~~\cite{Dress1955, Chen2006}. This makes it difficult to compare theoretical results with experimental observations.

In semiclassical theory, the lateral motion of spin would be attributed to the force exerted on the spin. From this point of view, the intrinsic force due to the k-linear spin-orbit coupling was found to be related to the conventional definition of spin current~\cite{Shen2005}. In the presence of disorder, the spin force in the Rashba–Dresselhaus system vanishes by accounting for the vertex correction~\cite{Chen2009}. However, in the Drude model regime, an effective Lorentz force can be generated by the crystal field~\cite{Eugene2007}, and it was shown to be related to the spin-Hall effect. Importantly, the spin-Hall conductivity generated by the effective Lorentz force is proportional to the charge conductance, and the resulting spin current fits experimental values observed for metals~\cite{Eugene2007}, which sheds light on the evaluation of the observable spin motion in the Drude model regime. It also worths noting that the semiclassical analysis of spin dynamics in the intrinsic Luttinger system was investigated by using adiabatic approximation~\cite{Jiang2005}.

In this study, we attempt to develop spin dynamics for two-dimensional spin-orbit coupled systems in the Drude model regime. The spin dynamics is treated as semiclassical motion governed by the Heisenberg equation of motion. The spin precession is ascribed to the rotation around the successive change in the effective magnetic field. To preserve the magnitude of the spin, the spin precession up to the first order of the electric field is comprised of Larmor and non-Larmor precessions. The former is a periodic motion and the latter is an exponential decay function. The spin-Hall conductivity (SHC) was also calculated by using the two components.

We find that in the regime $t/\tau\ll1$, the non-Larmor SHC is exactly equal to the result obtained from the Kubo formula. The formalism of the Kubo formula does not consider the Larmor precession. At finite $\tau$, we find that both the Larmor SHC and non-Larmor SHC vanish when the elapsed time is much larger than the relaxation time $\tau$. In the strong disorder limit (finite $\tau$ and vanishing Rashba coupling), both the Larmor and non-Larmor SHCs vanish independently over a finite time. In the intrinsic limit (infinite $\tau$ compared to the Larmor frequency), the Larmor SHC exactly cancels the non-Larmor SHC in the Kubo formula regime ($t/\tau\ll1$), which solves the problem of determining the universal constant of the SHC in previous studies. We also calculated the time-averaged spin-Hall conductivity for the k-cubic Rashba system~\cite{Wund2005}. The result is close to the experimental value. We note that the time-averaged quantum Hall conductivity is also investigated in~\cite{Liu2019}.

The remainder of this paper is organized as follows. In \secref{Drude}, the influence of the momentum relaxation time on the spin-orbit coupled system is investigated. The spin motion up to the first order of the electric field is composed of Larmor and non-Larmor precessions. The boundary condition is also determined by using preservation of the spin magnitude. The analytical solutions for the Larmor and non-Larmor precessions are presented in \secref{Larmor}. The relationship between the non-Larmor precession and the Kubo formula is discussed. In \secref{LNSHC}, the spin-Hall conductivities for both Larmor and non-Larmor precessions are calculated. In the intrinsic Rashba system, the resulting spin-Hall conductivity drops to zero when the Rashba coupling drops to zero. We also calculate the Larmor and non-Larmor spin-Hall conductivities for the k-cubic Rashba system. The resulting spin-Hall conductivity is shown to be close to the experimental value. Finally, the conclusion is presented in \secref{Concl}.

\section{2D Hamiltonian and Drude Model}\label{Drude}

The two-dimensional spin-orbit coupled Hamiltonian under consideration is given by
\begin{equation}\label{H0}
 H_0=\frac{\hbar^2k^2}{2m}+\frac{\hbar}{2}\Bsigma\cdot\BOmega_0,
\end{equation}
where $\Bsigma=(\sigma_x,\sigma_y,0)$ with $\Bsigma$ been Pauli matrices and $\BOmega_0=(2d_x/\hbar,2d_y/\hbar,0)$ is the Larmor frequency. The terms $d_x$ and $d_y$ depend on the spin-orbit interaction and two-dimensional momentum $\mk=(k_x,k_y)$. There is no term with $d_z$, which implies that the spin is forced to lie along the plane. For example, in the Rashba–Dresselhaus system, we have $d_x=\alpha k_y+\beta k_x$ and $d_y=-\alpha k_x-\beta k_y$. For the k-cubic Rashba-Dresselhaus system~\cite{Bulaev2005}, the Hamiltonian is given by $H_0=\epsilon_k+i\alpha(\sigma_+k_-^3-\sigma_-k_+^3)-\beta(\sigma_+k_-k_+k_-+\sigma_-k_+k_-k_+)$.
In this case, $\sigma_{\pm}=(\sigma_x\pm i\sigma_y)/2$ and $k_{\pm}=k_x\pm ik_y$. The corresponding $d_x$ and $d_y$ are defined as
$d_x=[\alpha\sin(3\phi)+\beta\cos(\phi)]k^3$, and $d_y=[-\alpha\cos(3\phi)+\beta\sin(\phi)]k^3$,
where $\phi=\tan^{-1}(k_y/k_x)$. The eigenenergies and eigenvectors are obtained by solving $H_0|n\mk\rangle=E_{n\mk}|n\mk\rangle$, which is given by
\begin{equation}\label{EigenE}
E_{nk}=\epsilon_k-nd,
\end{equation}
where $d=\sqrt{d_x^2+d_y^2}$ and $n=\pm$ is the energy band index. In the two-dimensional spin-orbit coupled systems, it is easy to show that the spin z component in the unperturbed Hamiltonian \eqnref{H0} is zero, i.e., $\langle{n\mk}|\sigma_z|{n\mk}\rangle$=0.

We consider a simple model in which the momentum and the Fermi-Dirac
distribution eventually approach a finite displacement through
the collision between electrons and impurities. In this paper, an electric charge is denoted as $-q$ and $q>0$. The equation of motion of
momentum in the presence of a constant electric field is given by
\begin{equation}\label{Drude-EOM}
\hbar\frac{d\mk_t}{dt}=-q\mathbf{E}+\frac{\hbar(\mk-\mk_t)}{\tau},
\end{equation}
where $\mk$ is the electron's momentum before the electric field
	is switched on, and $\tau$ is the momentum relaxation time. \eqnref{Drude-EOM} can be exactly solved, and the result is given
by~\cite{Jone1985}
\begin{equation}\label{Drude-EOMsol}
\mk_t=\mk-\frac{q\mathbf{E}}{\hbar}t^{\ast},
\end{equation}
where
\begin{equation}\label{timescale}
t^{\ast}=\tau\left(1-e^{-t/\tau}\right).
\end{equation}
Note that $t^{\ast}(0)=0$ and $t^{\ast}(\infty)=\tau$. When $\tau$ is very large (the clean limit), we have
$e^{-t/\tau}=1-t/\tau+t^2/2\tau^2+\cdots$ and $t^{\ast}(t/\tau\ll1)\rightarrow t$. This implies that
$\mk_t=\mk-q\mathbf{E}t/\hbar$ if $t\ll\tau$,
which is the same result as when impurities are not present. We refer to the limit $t/\tau\ll1$ as the clean limit.
	In this limit $t/\tau\ll1$, equilibrium cannot be achieved.

As time $t$ is much larger than $\tau$, the momentum has a finite
displacement of $-\tau q\mathbf{E}/\hbar$. The Fermi-Dirac
distribution $f_{n\mk}$ in $\mk$-space is also displaced by the
electric field such that ($t\gg\tau$)
\begin{equation}\label{disf}
\begin{split}
f_{n\mk_t}&=f_{n\mk-\tau q\mathbf{E}/\hbar}\\
&=f_{n\mk}-\tau\frac{q\mathbf{E}}{\hbar}\cdot\frac{\partial
	f_{n\mk}}{\partial\mk}+\cdots.
\end{split}
\end{equation}
 The first order correction to the distribution function is proportional to the relaxation time which gives the Drude's result of charge conductivity~\cite{Jone1985}. Furthermore, we note that the quantum kinetic equations including spin-orbit coupling has been investigated in Ref. \cite{Bryksin2006}, in which the first order correction of distribution function is related to the spin density matrix. However, in this paper, the first order correction of distribution function is irrelevant in the linear response regime. This is because the spin precession around the effective magnetic field is also caused by the applied electric field and moreover the unperturbed spin z component is zero [see Eqs. (\ref{TDSHC}) and (\ref{TDSHC1})].

The Heisenberg equation of motion of spin $\ms^0\equiv\langle{n\mk}|e^{iH_0t/\hbar}\Bsigma e^{-iH_0t/\hbar}|{n\mk}\rangle$ is given by
\begin{equation}\label{Leq}
\frac{\partial}{\partial t}\ms^0=\BOmega_0\times\ms^0,
\end{equation}
where $\BOmega_0$ depends on momentum $\mk$. In the presence of an applied electric field, the change in momentum would lead to the change
in the effective magnetic field and the spin orientation varies due to the precession. Semiclassically, we assume that the effect of disorder and the applied electric field are included in the effective magnetic field and the spin dynamics is still governed by the Heisenberg equation of motion. % (see Appendix~\ref{HEOM})
%The disorder and applied electric field that changes the momentum are included in the effective magnetic field, 
Thus, the resulting Hamiltonian can be written as
\begin{equation}\label{App-H-1}
H(t)=\frac{\hbar^2\mk_t^2}{2m}+\frac{\hbar}{2}\Bsigma\cdot\BOmega(t),
\end{equation}
where $\mk_t$ is given by \eqnref{Drude-EOMsol}, 
$\BOmega(t)=2\mathbf{d}(t)/\hbar$ and $\mathbf{d}(t)=(d_x(\mk_t),d_y(\mk_t),0)$. We require that the Hamiltonian \eqnref{App-H-1} satisfies the Schrodinger equation
\begin{equation}
i\hbar\frac{\partial}{\partial t}|\psi(t)\rangle=H(t)|\psi(t)\rangle.
\end{equation}
The time-dependent spin $\ms(t)$ is defined as $\ms(t)\equiv\langle\psi(t)|\Bsigma|\psi(t)\rangle$ which is the averaged spin in the presence of disorder and an electric field. Apply the first derivative of time to $\ms(t)$, we have
\begin{equation}\label{App-H-2}
\begin{split}
\frac{\partial}{\partial t}\ms(t)&=\langle\frac{\partial}{\partial t}\psi(t)|\Bsigma|\psi(t)\rangle+\langle\psi(t)|\Bsigma|\frac{\partial}{\partial t}\psi(t)\rangle\\
&=\langle\psi(t)|\frac{1}{i\hbar}[\Bsigma,H(t)]|\psi(t)\rangle,
\end{split}
\end{equation}
where \eqnref{App-H-1} was used. By using the commutator for Pauli matrices $[\sigma_i,\sigma_j]=2i\epsilon_{ijk}\sigma_k$, we obtain the Heisenberg equation of motion for the averaged spin,
\begin{equation}\label{Hspin}
\frac{\partial}{\partial t}\ms(t)=\BOmega(t)\times\ms(t),
\end{equation}
where $[\Bsigma,\mk_t^2]=0$ was used. We emphasize that the equation of motion for momentum is derived by using semiclassical Drude model (Eq.\eqref{Drude-EOMsol}), not the expectation value of $\mk$ with respect to the state $|\psi(t)\rangle$, i.e., $\mk_t\neq\langle\psi(t)|\mk|\psi(t)\rangle$.

% The time dependence of momentum $\mk_t$ is obtained by accounting for the semiclassical theory under consideration as shown in Eq. \eqnref{Drude-EOMsol}.%the following.

We now return to the momentum dependence of the effective magnetic field. The influence of the disorder in the presence of an electric field will lead to a change in the effective magnetic field, that is, $d_i(t)=d_i(\mk_t)$. Therefore, up to the linear order of the electric field,
\begin{equation}\label{dfield}
\begin{split}
d_i(\mk_t)&=d_i(k_a-\frac{qE_a}{\hbar}t^{\ast})\\
&=d_i-\frac{qE_a}{\hbar}t^{\ast}\frac{\partial
d_i}{\partial k_a}+o(E_a^2).
\end{split}
\end{equation}
Similar to the Fermi-Dirac distribution in the presence of disorder, when $t\rightarrow\infty$, the $\mathbf{d}$ vector changes its direction from $d_i$ to another fixed direction $d_i-(\tau eE_a/\hbar)(\partial d_i/\partial k_a)$. We define $\BOmega(t)=2\mathbf{d}(t)/\hbar=\BOmega_0+\BOmega'(t)$, where $\BOmega_0=(2d_x/\hbar,2d_y/\hbar,0)$ and
\begin{equation}\label{PLF}
\BOmega'(t)=-\frac{qE_a}{\hbar}t^{\ast}\frac{\partial
\BOmega_0}{\partial k_a},
\end{equation}
which also lies on the 2D plane.
The change in the effective magnetic field will cause the spin to precess around the new direction of the effective magnetic field. The spin’s z component will be non-zero at this time. The trajectory of $\mathbf{d}(t)$ in $d$-space is a straight line up to the first order of the electric field. The resulting angular velocity of $\mathbf{d}$ measured from the origin must vary with time. Therefore, we cannot use a rotating frame with fixed spin because the resulting Hamiltonian does not commute at different times. Instead, we use the perturbative method defined in Eq. (\ref{Hspin}).  The unitarity is broken and the magnitude of the spin is preserved only up to the first order of an electric field. Namely, we have
\begin{equation}
\ms(t)=\ms^0(t)+\ms'(t)+o(E_a^2),
\end{equation}
where $\ms^0$ and $\ms'$ correspond to the solutions without an electric field and with an electric field, respectively. Inserting this result into Eq. (\ref{Hspin}),
\begin{equation}\label{NLeq}
\frac{\partial}{\partial t}\ms'=\BOmega_0\times\ms'+\BOmega'\times\ms^0,
\end{equation}
where the following unperturbed Eq. (\ref{Leq}) was used.
The solution of Eq. (\ref{Leq}) (using $|\ms^0|=1$) is given by
\begin{equation}\label{LeqSol}
\begin{split}
&s^0_x=\cos\Theta\cos\theta_0-\sin\Theta\sin\theta_0\sin(\Omega_0t),\\
&s^0_y=\sin\Theta\cos\theta_0+\cos\Theta\sin\theta_0\sin(\Omega_0t),\\
&s^0_z=-\sin\theta_0\cos(\Omega_0t),
\end{split}
\end{equation}
where the angle $\theta_0$ is the angle between $\BOmega_0$ and $\ms^0$, and $\Theta$ is defined as $\tan\Theta=d_y/d_x$.
Eq. (\ref{Leq}) implies that $\ms^0$ behaves as the spin in the intrinsic system $H_0$, where the Larmor frequency is $\BOmega_0$.
The spin should be aligned with $\BOmega_0$ without applying an electric field. In this sense, we can assume that $\theta_0=0$, and we have
\begin{equation}\label{Leq1}
\begin{split}
&s^0_x=\cos\Theta=\frac{d_x}{d},\\
&s^0_y=\sin\Theta=\frac{d_y}{d},\\
&s^0_z=0.
\end{split}
\end{equation}
To obtain results for different energy bands, we can simply use the replacement $\ms_0\rightarrow(-n)\ms_0$ in Eq. (\ref{Leq1}), where $n=\pm$ is the band index. Substituting Eq. (\ref{Leq1}) with the appropriate band index into Eq. (\ref{NLeq}), we have ($\lambda_a\equiv qE_a$)
\begin{equation}\label{Drude-sigmaz}
\frac{\partial}{\partial
t}s_z'=-\frac{\lambda_at^{\ast}}{\hbar}n\Omega_0\Gamma^z_a
+s_y'\Omega_{0x}-s_x'\Omega_{0y},
\end{equation}
for the spin-z component, and
\begin{equation}\label{Drude-sigmaxandy}
\begin{split}
\frac{\partial}{\partial
t}s_x'&=\Omega_{0y}s_z',\\
\frac{\partial}{\partial t}s_y'&=-\Omega_{0x}s_z',
\end{split}
\end{equation}
for the spin x and y components, where $\Gamma^z_a$ is defined as
\begin{equation}\label{Gammaz}
\Gamma_a^z=\frac{1}{d^2}\left(d_x\frac{\partial d_y}{\partial k_a}-d_y\frac{\partial d_x}{\partial k_a}\right).
\end{equation}

We observe that the term $\BOmega'(t)\times\ms^0$ in Eq. (\ref{NLeq}) behaves as a torque such that
$\ms'$ not only has a Larmor rotation (periodic rotation) but also a non-Larmor rotation (non-periodic motion). Therefore, similar to the derivations in
a previous study, the spin will be composed of non-Larmor (N) and Larmor (L) precession terms.
\begin{equation}
\ms'=\BSigma^L+\BSigma^N
\end{equation}
The term $\BSigma^L$ is called the Larmor component and is governed by
\begin{equation}\label{BSigmaLeq}
\frac{\partial}{\partial t}\BSigma^L=\BOmega_0\times\BSigma^L.
\end{equation}
Eq. (\ref{BSigmaLeq}) is similar to Eq. (\ref{Leq}); however, $\BSigma^N$ contains the response of the electric field, and we expect that the direction of $\BSigma^L$ is not parallel to the static Larmor frequency $\BOmega_0$. On the other hand, the term $\BSigma^N$ is called the non-Larmor component, which satisfies
\begin{equation}\label{BSigmaNeq}
\frac{\partial}{\partial t}\BSigma^N=\BOmega_0\times\BSigma^N+\BOmega'\times\ms^0.
\end{equation}
It must be emphasized that Eqs. (\ref{BSigmaLeq}) and (\ref{BSigmaNeq}) hold only if the magnitude of spin $\ms(t)$ is valid up
to the second order of electric field. The magnitude of spin is given by $|\ms|^2=|\ms^0+\ms'+o(E_a^2)|^2=|\ms^0|^2+2\ms^0\cdot\ms'+o(E_a^2)$. To preserve the magnitude of spin up to the second order of $E_a$, we require that
\begin{equation}\label{sdotsp}
\ms^0\cdot\ms'=0
\end{equation}
at any time $t$. Furthermore because the time dependence of the Larmor component is different from that of non-Larmor components, Eq. (\ref{sdotsp}) implies that
\begin{equation}\label{sdotLN}
\begin{split}
&\ms^0\cdot\BSigma^L=0,\\
&\ms^0\cdot\BSigma^N=0.
\end{split}
\end{equation}
On the other hand, at $t=0$, the precession of spin begins from the in-plane and ends as an out-of-plane orientation. In this sense, the z component $s_z'(0)$ must be zero, although its rate of change is non-zero. This can be described by
\begin{equation}\label{szero}
\begin{split}
&s'_z(0)=0,\\
&\left(\frac{\partial s_z'}{\partial t}\right)_{t=0}=\Omega_0\left(\cos\Theta s_y'(0)-\sin\Theta s_x'(0)\right)\neq0,
\end{split}
\end{equation}
where Eq. (\ref{NLeq}) at $t=0$ was used. We note that a similar derivation was proposed in Ref.~\cite{Paul2018}, in which the effective magnetic field is replaced by a pseudo-magnetic field derived from the spin force. In the following section, we solve $\ms'$ by acquiring the boundary conditions for Eqs. (\ref{sdotLN}) and (\ref{szero}).

\section{Larmor and Non-Larmor precessions}\label{Larmor}
The solutions of
\eqnref{BSigmaLeq} would be the same as Eq. (\ref{LeqSol}). That is, we have
\begin{equation}\label{Ltheta}
\begin{split}
&\BSigma^L_x=\Sigma^L\left[\cos\Theta\cos\theta-\sin\Theta\sin\theta\sin(\Omega_0t)\right],\\
&\BSigma^L_y=\Sigma^L\left[\sin\Theta\cos\theta+\cos\Theta\sin\theta\sin(\Omega_0t)\right],\\
&\BSigma^L_z=\Sigma^L\left[-\sin\theta\cos(\Omega_0t)\right].
\end{split}
\end{equation}
However, the difference is that $\BSigma^L$ is in linear order with the applied electric field in the disordered system. The angle $\theta$ between $\BSigma^L$ and $\BOmega_0$ depends on the electric field and relaxation time $\tau$. First, by using the requirement in Eq. (\ref{sdotLN}), we have $\Sigma^L_x\cos\Theta+\Sigma^L_y\sin\Theta=0$, which implies that $\Sigma^L\cos\theta=0$. Because $\Sigma^L\neq0$, we obtain $\theta=\pi/2$, which means that $\BSigma^L$ is always perpendicular to $\BOmega_0$. Eq. (\ref{Ltheta}) then becomes
\begin{equation}\label{Ltheta1}
\begin{split}
&\BSigma^L_x=-\Sigma^L\sin\Theta\sin(\Omega_0t),\\
&\BSigma^L_y=\Sigma^L\cos\Theta\sin(\Omega_0t),\\
&\BSigma^L_z=-\Sigma^L\cos(\Omega_0t).
\end{split}
\end{equation}
Applying the time derivative to the z component of \eqnref{BSigmaNeq}, and noting that $dt^{\ast}/dt=e^{-t/\tau}$, we obtain
\begin{equation}\label{NL-1}
\frac{\partial^2}{\partial
t^2}\Sigma^N_{z}+\Omega_0^2\Sigma^N_{z}+\frac{n\lambda_a}{\hbar}\Omega_0\Gamma^z_ae^{-t/\tau}=0,
\end{equation}
where $\Omega_0=2d/\hbar$ was used. The solution of Eq. (\ref{NL-1}) is given by
\begin{equation}\label{SigmaNz}
\Sigma^N_{z}=A_ne^{-t/\tau},
\end{equation}
where $A_n$ is the dimensionless quantity
\begin{equation}\label{An}
A_n=-\frac{n\lambda_a}{\hbar}\Omega_0\Gamma^z_a\frac{1}{\frac{1}{\tau^2}+\Omega_0^2}.
\end{equation}
Eq. (\ref{SigmaNz}) plays an important role in the spin-Hall effect. We will return to this point in the next section. We now use Eq. (\ref{szero}), which is $\Sigma^N_z(0)+\Sigma^L_z(0)=0$. Thus, $A_n-\Sigma^L=0$, and the resulting z component of $\BSigma^L$ is given by
\begin{equation}\label{SigmaLz}
\Sigma^L_{z}=-A_n\cos(\Omega_0 t)
\end{equation}
and the x and y components of $\BSigma^L$ are given by
\begin{equation}\label{SigmaLxy}
\begin{split}
\Sigma^L_{x}=&-A_n\sin\Theta\sin(\Omega_0
t),\\
\Sigma^L_{y}=&A_n\cos\Theta\sin(\Omega_0
t).
\end{split}
\end{equation}

Substituting Eqs. (\ref{SigmaNz}) and (\ref{An}) into Eq. (\ref{BSigmaNeq}) for the x and y components, we obtain
\begin{equation}\label{SigmaNxy}
\begin{split}
\Sigma^N_{x}&=-A_n\Omega_0\tau\left(e^{-t/\tau}-\frac{\frac{1}{\tau^2}+\Omega_0^2}{\Omega_0^2}\right)\sin\Theta,\\
\Sigma^N_{y}&=A_n\Omega_0\tau\left(e^{-t/\tau}-\frac{\frac{1}{\tau^2}+\Omega_0^2}{\Omega_0^2}\right)\cos\Theta,
\end{split}
\end{equation}
Arbitrary time-independent constants $C_x$ and $C_y$ can be added to $\Sigma^N_x$ and $\Sigma^N_y$, respectively. The two constants must obey the condition $d_xC_y-d_yC_x=0$. Furthermore, the preserved magnitude of spin $|\ms(t)|=1+o(E_a^2)$ (see Eq. (\ref{sdotLN})) implies that $d_xC_x+d_yC_y=0$. The only solution is $C_x=C_y=0$ if $d\neq0$. On the other hand, it can be shown that Eqs. (\ref{SigmaNz}), (\ref{SigmaLz}), (\ref{SigmaLxy}), and (\ref{SigmaNxy}) satisfy the second boundary condition of Eq. (\ref{szero}).

When $t\rightarrow\infty$, the momentum has a constant shift and the effective magnetic field does not change its direction over time. The system achieves an equilibrium state. We find that the in-plane components $\Sigma^N_x$ and $\Sigma^N_{y}$ do not change with time, and
the non-Larmor spin z component $\Sigma^N_z$ vanishes. Only the $\BSigma^L$ Larmor components survive in the system. In the next section, we calculate the Larmor SHC and non-Larmor SHC in the Rashba system. We close this section by discussing the relationship between our results and the Kubo formula in the intrinsic system.

If the energy gap is much larger than the broadening of the energy band due to the disorder, the system is said to be within the intrinsic limit, that is, $2d\gg\hbar/\tau$.  This implies that the intrinsic limit is given by $\Omega_0\tau\gg1$ when $\mk\neq0$. When $\tau$ is finite, the intrinsic limit implies that the system is in the strong spin-orbit coupling regime and cannot be zero unless $\tau\rightarrow\infty$. Therefore, the intrinsic limit leads to the result that
the system displays a physical evolution only for a very short time compared to a finite $\tau$. In the following, we will demonstrate that at the intrinsic limit (and clean limit), the non-Larmor spin z component $\Sigma^N_z$ is exactly equal to the result obtained from the Kubo formula.

If we take the intrinsic limit $\Omega_0\tau\gg1$, then $\Sigma^N_i$ is
proportional to $(e^{-t/\tau}-1)\Omega_0\tau$. Then, taking the clean limit $t/\tau\ll1$, we have $(e^{-t/\tau}-1)\Omega_0\tau\approx-\Omega_0t$. Therefore, considering the
limit $\Omega_0\tau\gg1$ and then $t/\tau\ll 1$, we have
\begin{equation}\label{Drude-Climit}
\begin{split}
\Sigma^N_{x}&\approx+A_n\Omega_0t\sin\Theta,\\
\Sigma^N_{y}&\approx-A_n\Omega_0t\cos\Theta.\\
A_n&\approx-\frac{n\lambda_a}{\hbar\Omega_0}\Gamma_a^z.
\end{split}
\end{equation}
Note that in the limit $\Omega_0\tau\gg1$ with $t/\tau\ll1$, the non-Larmor spin x and y components exhibit no physical growth in time. Furthermore, Eq. (\ref{Drude-Climit}) can be exactly derived from the Heisenberg equation of motion~\cite{Chen2019}. If we substitute Eq. (\ref{Drude-Climit}) back into Eq. (\ref{BSigmaNeq}), we obtain
\begin{equation}\label{SigmazN}
\begin{split}
&\frac{\partial\Sigma^N_z}{\partial t}\approx0,\\
&\Sigma^N_{z}=A_n=-\frac{n\lambda_a}{\hbar\Omega_0}\Gamma^z_a.
\end{split}
\end{equation}
Interestingly, because of the linear time dependence of the spin x and y components, the response
of the non-Larmor spin z component is a constant only for a very short time. It has been shown that Eq. (\ref{SigmazN}) is exactly the same as the result obtained from the Kubo formula in the intrinsic case~\cite{Chen2014,Chen2019}. The spin-Hall current from the Kubo formula using the result in Eq. (\ref{SigmazN}) is a universal constant in the Rashba system. The universal constant also leads to the problem that the spin x and y components have no physical growth with time, as has been demonstrated in Refs~~\cite{Chen2006,Chen2019,Dim2005,Chalaev2005}. Nevertheless, as discussed in the above results, the regime of the Kubo formula in the intrinsic spin-orbit coupled system is valid only for a very short time limit. In particular, the short time limit corresponds to the time when the spin adiabatically aligns its orientation with the effective magnetic field. Consequently, the change in the magnitude of the effective magnetic field (and thus the spin in that direction) yields a non-zero spin z component in order to preserve the spin magnitude~\cite{Sinova2004}.

\section{time-dependent spin-Hall effect}\label{LNSHC}

The spin-Hall current deduced from the spin-dynamics would be semiclassically obtained by using the conventional definition of the spin current, which is the simple multiplication of spin $(J\hbar/2)s_z$ and the lateral velocity $\hbar k_x/m$ in which the electric field is applied in y-direction, where $J=1$ for the Rashba–Dresselhaus system and $J=3$ for the k-cubic Rashba system. The time-dependent spin-Hall current is given by
\begin{equation}\label{TDSHC}
\begin{split}
\mathcal{J}^z_x&=\sum_{n\mk}f_{n\mk_t}\frac{J\hbar}{2}s_z\frac{\hbar k_x}{m}\\
&=\sum_{n\mk}(f_{n\mk}-t^{\ast}\frac{q E_y}{\hbar}\cdot\frac{\partial
f_{n\mk}}{\partial k_y})\frac{J\hbar}{2}(s^0_z+s_z')\frac{\hbar k_x}{m}.\\
\end{split}
\end{equation}
We note that $s^0_z=0$ (see \eqnref{Leq1}) in the present systems under consideration. Up to the first order of the applied electric field, the time-dependent spin-Hall current \eqnref{TDSHC} becomes
\begin{equation}\label{TDSHC1}
\mathcal{J}^z_x=\sum_{n\mk}f_{n\mk}\frac{J\hbar}{2}s'_z\frac{\hbar k_x}{m}=\sigma^z_{xy}E_y,
\end{equation}
The linear response of the spin z component $s'_z$ is given by $s_z'=\Sigma^L_z(t)+\Sigma^N_z(t)$ [see Eqs. (\ref{SigmaNz}) and (\ref{SigmaLz})]. The spin-Hall conductivity $\sigma^z_{xy}$ can now be written as the Larmor component $\sigma^L_{xy}$ and non-Larmor component $\sigma^N_{xy}$, that is,
\begin{equation}\label{SHC}
\sigma^z_{xy}=\sigma^{L}_{xy}+\sigma^{N}_{xy}.
\end{equation}

\begin{figure}
\begin{center}
\includegraphics[width=8cm,height=6cm]{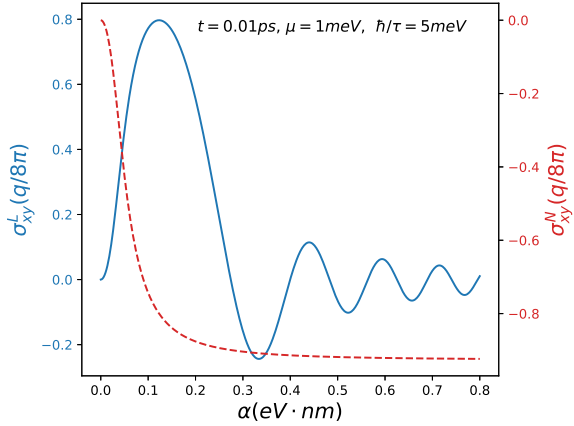}
\end{center}
\caption{(color online) Numerical results of Larmor ($\sigma^L_{xy}$) and non-Larmor ($\sigma^N_{xy}$) spin-Hall conductivities in the presence of disorder with finite $\tau$. When the Rashba coupling $\alpha$ approaches zero (corresponding to a strong disorder), both SHCs drop to zero independently.}\label{FigR1}
\end{figure}

\begin{figure}
\begin{center}
\includegraphics[width=8cm,height=6cm]{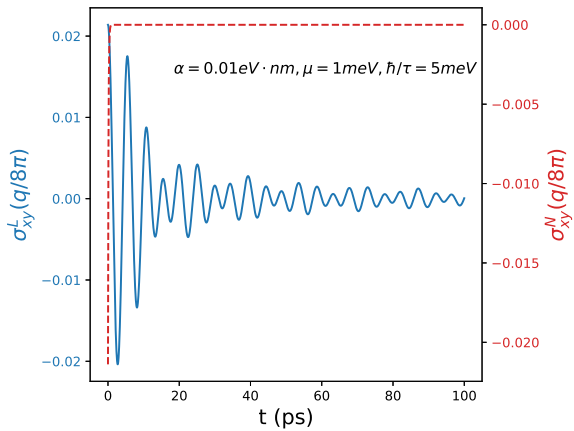}
\end{center}
\caption{(color online) Numerical values of $\sigma^L_{xy}$ and $\sigma^N_{xy}$ for the Rashba system with finite $\tau$. As time increases beyond $\tau$, $\sigma^N_{xy}$ vanishes, while $\sigma^{L}_{xy}$ survives but decays over time.}\label{FigR2}
\end{figure}

\begin{figure}
\begin{center}
\includegraphics[width=8cm,height=6cm]{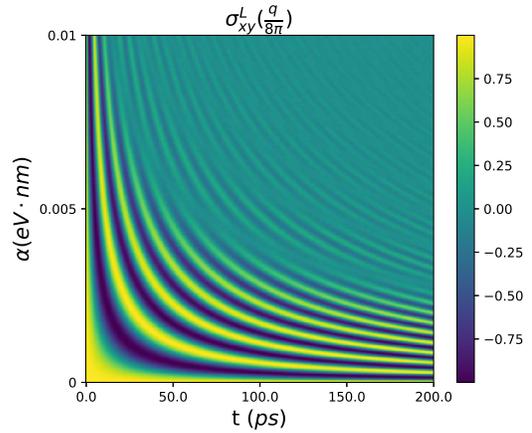}
\end{center}
\caption{(color online)Numerical values of the Larmor spin-Hall conductivity ($\sigma^L_{xy}$) for the Rashba system in the intrinsic limit. At finite time $t$, when the Rashba coupling drops to zero, $\sigma^L_{xy}$ approaches the universal constant but is opposite to $\sigma^N_{xy}$ in sign.}\label{FigR3}
\end{figure}

\subsection{k-linear Rashba system}

Considering the k-linear Rashba system, it can be shown that for the Larmor SHC,
\begin{equation}\label{L-SHC}
\sigma^L_{xy}=\frac{\alpha q}{4\pi m}\int_{k^-_F}^{k^+_F}dk\frac{k^2\cos(2\alpha kt/\hbar)}{\frac{1}{\tau^2}+(\frac{2\alpha k}{\hbar})^2}.
\end{equation}
For the non-Larmor component,
\begin{equation}\label{N-SHC}
\sigma^N_{xy}=-\frac{\alpha q}{4\pi m}e^{-t/\tau}\int_{k^-_F}^{k^+_F}dk\frac{k^2}{\frac{1}{\tau^2}+(\frac{2\alpha k}{\hbar})^2},
\end{equation}
where the Fermi momenta at the two bands $k_F^{\pm}$ are given by
\begin{equation}\label{FermiM}
\frac{k_F^{\pm}}{k_0}=\pm1+\sqrt{1+\frac{\mu}{\mathcal{E}_R}},~k_0\equiv\frac{m\alpha}{\hbar^2},
\end{equation}
where $\mu$ is the Fermi energy. We define the Rashba energy $\mathcal{E}_R$ as $\mathcal{E}_R=\hbar^2k_0^2/2m=m\alpha^2/2\hbar^2$. Equation (\ref{N-SHC}) is then exactly the same as the spin-Hall conductivity obtained from the Kubo formula in the presence of disorder~\cite{Loss2004}, except for the time dependence $e^{-t/\tau}$. We find that in the presence of a strong disorder (finite $\tau$ and vanishing Rashba coupling), the Larmor and non-Larmor SHCs vanish independently, as shown in Fig. \ref{FigR1}. Furthermore, when $\tau$ is finite, the time evolution of the Larmor and non-Larmor SHCs is shown in Fig. \ref{FigR2}. We find that the non-Larmor SHC approaches zero faster than the Larmor SHC. The Larmor SHC survives but decays over time.

Consider the intrinsic limit $\tau\rightarrow\infty$ in the k-linear Rashba system, where the time $t$ can be considered finite in this limit. It can be shown that
\begin{equation}\label{R-SHC3}
\begin{split}
&\sigma^L_{xy}=\frac{q}{8\pi}\frac{\hbar}{4\alpha k_0
t}\left[\sin\left(\frac{2\alpha
k_F^+}{\hbar}t\right)-\sin\left(\frac{2\alpha
k_F^-}{\hbar}t\right)\right],\\
&\sigma^N_{xy}=-\frac{q}{8\pi},
\end{split}
\end{equation}
where $e^{-t/\tau}\approx1$ for $\sigma^N_{xy}$. The universal constant in $\sigma^N_{xy}$
is exactly the same as the result obtained from the Kubo formula. When the two bands are occupied, $k_F^+$ is not equal to
$k_F^-$, and the oscillating term in $\sigma^L_{yx}$ is always finite. As time progresses, $\sigma^L_{xy}\rightarrow0$, which is the same as the result with finite $\tau$. When the Rashba coupling is very small (for finite $t$),
\begin{equation}\label{sk-sk}
\begin{split}
&\sin\left(\frac{2\alpha k_F^+t}{\hbar}\right)-\sin\left(\frac{2\alpha k_F^-t}{\hbar}\right)\\
&=\frac{2\alpha t}{\hbar}\left(k_F^+-k_F^-\right)+o(\alpha^6t^3)\\
&=\frac{4\alpha k_0t}{\hbar}+o(\alpha^6t^3),
\end{split}
\end{equation}
where Eq. (\ref{FermiM}) was used. We also note that $k_F^{\pm}$ is the order of $\alpha$. Substituting Eq. (\ref{sk-sk}) into Eq. (\ref{R-SHC3}),
\begin{equation}
\sigma^z_{xy}=\sigma^L_{xy}+\sigma^N_{xy}=\frac{q}{8\pi}-\frac{q}{8\pi}=0.
\end{equation}
Therefore, we find that when the Larmor motion is considered, we can solve the problem caused by the Kubo formula. The numerical values for the Larmor SHC in Eq. (\ref{R-SHC3}) are shown in Fig.~\ref{FigR3}. We find that at finite $t$, vanishingly small Rashba coupling indeed exhibits a region where $\sigma^L_{xy}\rightarrow q/8\pi$.

In short, the spin dynamics preserve the spin magnitude up to the first order of the electric field. In this perturbation method, spin motion is composed of Larmor and non-Larmor precessions. The Larmor SHC does not have the same time-dependent function as the non-Larmor SHC, and thus, in general, they cannot cancel each other. Furthermore, in the intrinsic system, the spin-Hall conductivity from the Kubo formula is equal to the non-Larmor SHC over a short time as compared to $\tau$. In the Rashba system, the spin-Hall conductivity is generally not a universal constant, and we have demonstrated that when Rashba coupling vanishes, the spin-Hall conductivity also vanishes.

\begin{figure}
\begin{center}
\includegraphics[scale=0.5]{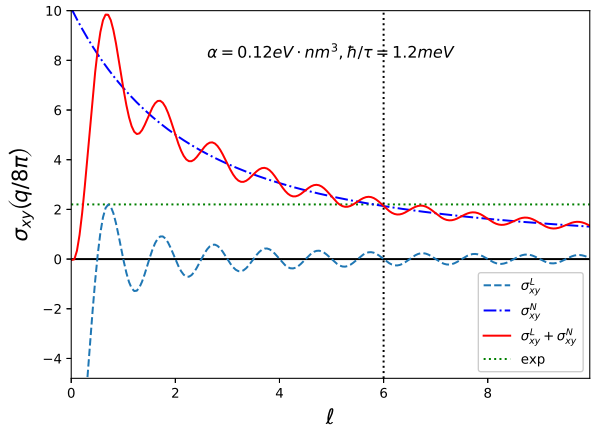}
\end{center}
\caption{(color online) Numerical values of the time-averaged spin-Hall conductivities vs. the parameter $\ell$ for the k-cubic Rashba system. The experimental value $2.2(e/8\pi)$ (green dotted line) corresponds to the value $\ell\approx6$ (the vertical dotted line). The Larmor/non-Larmor SHC is shown by blue dashed / dash-dotted line. The total SHC is shown by red solid line.}\label{FigR4}
\end{figure}

\subsection{k-cubic Rashba system}
We close this section by calculating the spin-Hall conductivity in the k-cubic Rashba system, in which the experimental value is $2.2(q/8\pi)$ in the clean limit~\cite{Wund2005}. In the intrinsic two-dimensional k-cubic Rashba system, the SHC is shown to be $9(q/8\pi)$ when the length scale $\hbar^2/2m\alpha$ is much larger than $\sqrt{4\pi n_h}$ ~\cite{Loss2005}, where $n_h$ is the 2D hole density. The reduction of the SHC in the intrinsic k-cubic Rashba system has been investigated by many authors. In Ref.~\cite{Bern2005}, the SHC is ascribed to the finite thickness of the quantum well, and the conventional definition of the spin current yields a value of $1.9(q/8\pi)$. By further taking into account the spin-torque dipole current in the quantum well with a finite thickness,  the authors in Ref.~\cite{Zhang2008} demonstrated that the SHC resulting from the SZXN spin current yields a value of $1.2(q/8\pi)$, which is also very close to the experimental value.

However, for the above two cases, the structure inversion asymmetric (SIA) Rashba term is neglected, although experiments were performed under strong SIA Rashba coupling. As indicated in Ref.~\cite{Zhang2008}, the inclusion of the SIA Rashba term would lead to a value of $-10(q/8\pi)$, which is higher than the experimental value, and the sign of SHC changes. We also note that the spin-Hall effect would be due to the edge spin accumulation~\cite{Nomura2005} and the magnetization at the edge of the sample would be in non-equilibrium~\cite{Bleibaum2006}. To compare the experimental values, we used the time-averaged formula for the time-dependent parts~\cite{tauavg},
\begin{equation}\label{TAterm}
\overline{f(t)}=\frac{\Omega_0}{2\pi\ell}\int_0^{2\pi\ell/\Omega_0}f(t)dt.
\end{equation}
That is, the system's response should have vanishingly small Larmor SHC in average. Equation~(\ref{TAterm}) implies that the time-averaged quantity is a function of $k$. This is because each $k$ point has its own Larmor frequency.
Nonetheless, since the spins do not interact with each other, there is no physical constraint demanding that the parameter should depend on k.  Each k point should be averaged by the same period.
The quantity $\ell$ is the parameter that can be tuned to fit the experimental value, which is the effective number of rotations from the initial spin state back to the initial spin state. ecause the wave function of spin should rotate $4\pi$ in order to return to its original wave function~\cite{Sak2017}, and we expect $\ell=2$ for k-linear spin-orbit coupled system. For k-cubic Rashba system, the rotational symmetry in momentum is three-fold in the wave function, and we expect $\ell=2\times 3=6$ (see Appendix~\ref{ENR}).

By using Eq. (\ref{TAterm}), the Larmor and non-Larmor SHCs can be written as (the detail is shown in Appendix~\ref{kcubicapp})
\begin{widetext}
\begin{equation}\label{k3Rashba}
\begin{split}
&\overline{\sigma^L_{xy}}=-\frac{9q}{8\pi}\frac{\hbar^2}{m}\frac{\sin(2\pi\ell)}{2\pi\ell}\int_{k_F^-}^{k_F^+}dk\frac{2\alpha k^4}{\left(\frac{\hbar}{\tau}\right)^2+\left(2\alpha k^3\right)^2},\\
&\overline{\sigma^N_{xy}}=\frac{9q}{8\pi}\frac{\hbar^2}{m}\frac{1}{2\pi\ell}\int_{k_F^-}^{k_F^+}dk\frac{2\alpha k^4}{\left(\frac{\hbar}{\tau}\right)^2+\left(2\alpha k^3\right)^2}\frac{2\alpha k^3}{(\hbar/\tau)}\left(1-e^{-2\pi\ell(\hbar/\tau)/2\alpha k^3}\right),
\end{split}
\end{equation}
\end{widetext}
The experimental value of the spin splitting is approximately $10^{-2}eV$ at the Fermi momentum: $k_F\approx0.35nm^{-1}$~\cite{Wund2005}. The estimated Rashba coupling is approximately $0.12 eV nm^{-3}$. The quasi-particle lifetime broadening is $\hbar/\tau\approx1.2\times10^{-3}eV$. The experimental value of the hole mass is $m=0.27m_0$. On the other hand, according to the observation of SHC in the clean two-dimensional hole gas \cite{Wund2005}, the hole concentration ($n_h$) is $n_h=2\times 10^{12}cm^{-2}$. The Fermi wave vectors can be extracted according to the equations \cite{Loss2005}
\begin{eqnarray}
k_F^{\pm}&=&\mp\frac{1}{2}\frac{\hbar^2}{2m\alpha}
	\left[
		1-\sqrt{1-\left(\frac{2m\alpha}{\hbar^2}\right)^2
			4\pi n_h}	
	\right]\\ \nonumber
	&+&
	\sqrt{
		\frac{-1}{2}\left(\frac{\hbar^2}{2m\alpha}\right)^2
		\left[1-\sqrt{1-\left(\frac{2m\alpha}{\hbar}\right)^2 4\pi n_h}\right] + 3\pi n_h
	}.
\end{eqnarray}
The numerical values of the Fermi wave vectors are $k_F^{+}=0.41nm^{-1}$ and $k_F^{-}=0.29nm^{-1}$. The numerical results of Eq. (\ref{k3Rashba}) are showed in Fig.~\ref{FigR4}.We find that the total SHC oscillates and follows non-Lamor SHC which is attributed to the adiabatic evolution of $\Sigma^N_z$. The experimental value $2.2(q/8\pi)$ corresponds to the parameter value $\ell\approx6$ as expected. The contribution from Larmor SHC is vanishingly small. By using $\ell=6$, we have $\overline{\sigma^N_{xy}}=2.1(q/8\pi)$.

\section{Conclusion}\label{Concl}
We solved the spin dynamics of two-dimensional spin-orbit coupled systems in the Drude model regime, in which the momentum relaxation time is taken into account. By considering the change in the effective magnetic field induced
by the applied electric field, the spin will precess around the successive change of the effective magnetic field. The spin dynamics were investigated
by using the perturbation method. The spin response to the linear order of the electric field is found to be composed of Larmor and non-Larmor precessions. The Larmor motion is the precession occurring around the unperturbed effective magnetic field, and the non-Larmor motion of spin is due to the extra torque induced by the electric field. The time-dependent spin-Hall conductivity (SHC) of the Rashba system was then calculated. In the presence of disorder, when time grows larger than the relaxation time, we  found that both the Larmor and non-Larmor SHCs drop to zero. However, the non-Larmor SHC decays faster than the Larmor SHC. This occurs because the presence of the disorder forces the system to achieve equilibrium, and the effective magnetic field eventually does not change over time. On the other hand, we found that the non-Larmor motion in the short time limit is exactly equal to the result obtained from the Kubo formula in the intrinsic case. In the intrinsic Rashba system, we found that the spin-Hall conductivity is not a universal constant. Furthermore, the spin-Hall conductivity vanishes when the Rashba coupling vanishes. We also calculated the Larmor and non-Larmor SHCs for the k-cubic Rahsba system. To meaningfully compare our results with the experimental value, we used the time-averaged spin-Hall conductivity. By comparing the calculated results to the experimental results, we found that the $6\times2\pi$ rotation for the spin wave function and the preservation of the magnitude of spin lead to the experimental result.

\begin{acknowledgments}
 T.-W.Chen would like to thank D.-W. Chiou for valuable discussions. This work was supported by the Ministry of Science and Technology of Taiwan under Grants No. 108-2112-M-110-009 and 109-2112-M-110-006, and 108-2112-M-004-002-MY2.
\end{acknowledgments}

\appendix

%\section{Heisenberg equation of motion}\label{HEOM}

\section{effective number of rotations}\label{ENR}
The parameter $\ell$ is the effective number of rotations of the spin to get back to the same initial state. For the unperturbed k-cubic Rashba model,
\begin{equation}\label{Hk3}
\begin{split}
H_0&=\frac{\hbar^2k^2}{2m}+i\alpha(k_-^3\sigma_+-k_+^3\sigma_-)\\
&=\left(\begin{array}{cc}
\epsilon_k&i\alpha k^3e^{-3i\phi}\\
-i\alpha k^3e^{3i\phi}&\epsilon_k
\end{array}\right)
\end{split}
\end{equation}
where $\epsilon_k=\hbar^2k^2/2m$, $\sigma_{\pm}=(\sigma_x\pm i\sigma_y)/2$, $k_{\pm}=k_x\pm ik_y$ and $\phi=\tan^{-1}(k_y/k_x)$. The unperturbed wave function $|\psi_0(t)\rangle$ is assumed to evolve according to the Schrodinger equation
\begin{equation}
i\hbar\frac{\partial}{\partial t}|\psi_0(t)\rangle=H_0|\psi_0(t)\rangle
\end{equation}
The eigenstates of \eqnref{Hk3} obeying $H_0|\pm\rangle=E_{\pm}|\pm\rangle$ are given by
\begin{eqnarray}
|{\pm\mk}\rangle=\frac{1}{\sqrt{2}}
\begin{pmatrix}
&\pm e^{3i\phi}& \\
&i&
\end{pmatrix},
\end{eqnarray}
and the corresponding eigenenergies are $E_{\pm\mk}=\epsilon_k\mp\alpha k^3$, where $\epsilon_k=\hbar^2k^2/2m$.
For the initial state $|\psi_i\rangle$ with a wave vector $k$, after time $t$, the time evolution of the state is given by
\begin{equation}
\begin{split}
&|\psi_0(t)\rangle\\
&=e^{-iH_0t/\hbar}|\psi_i\rangle\\
&=e^{-i\epsilon_kt/\hbar}\left(e^{+i\Omega_0 t/2}|{+\mk}\rangle\langle{+\mk}|\psi_i\rangle+e^{-i\Omega_0 t/2}|{-\mk}\rangle\langle{-\mk}|\psi_i\rangle\right),
\end{split}
\end{equation}
where $\Omega_0=2\alpha k^3/\hbar$ is the Larmor frequency. For a latter time $t+2\pi\ell/\Omega_0$, the state becomes
\begin{equation}
\begin{split}
&|\psi_0(t+\frac{2\pi\ell}{\Omega_0})\rangle=e^{-i\epsilon_k(t+2\pi\ell/\Omega_0)/\hbar}\\
&\times\left(e^{i\Omega_0 t/2}|{+\mk(\ell)}\rangle\langle
{+\mk}|\psi_i\rangle+e^{-i\Omega_0 t/2}|{-\mk(\ell)}\rangle\langle{-\mk}|\psi_i\rangle\right),
\end{split}
\end{equation}
where
\begin{eqnarray}
|{\pm\mk(\ell)}\rangle=
\frac{1}{\sqrt{2}}
\begin{pmatrix}
&\pm e^{3i(\phi\pm \pi\ell/3)}& \\
&ie^{\pm i\pi\ell}&
\end{pmatrix}.
\end{eqnarray}
For the spinor $|\psi_0(t+\frac{2\pi\ell}{\Omega_0})\rangle$ to get back to the same state as $|\psi_0(t)\rangle$ with the dynamical phase $\exp(i\epsilon_k(t+2\pi\ell/\Omega_0)t/\hbar)$, it is required that $\ell$ to be an even integer and a multiple of $3$. Thus, $\ell$ is given by the least common multiple of $2$ and $3$, which is $6$. The fitting parameter $\ell$ reflects the three-fold symmetry of the Hamiltonian and the eigenstates.

\section{SHC for k-cubic Rashba system}\label{kcubicapp}
The unperturbed Hamiltonian for the k-cubic Rashba system is given by \eqnref{Hk3}
which can be written as
\begin{equation}
H_0=\frac{\hbar^2k^2}{2m}+\sigma_xd_x+\sigma_yd_y.
\end{equation}
The angle $\phi$ is defined as $\tan\phi=k_y/k_x$ and $d_x$ and $d_y$ can be written as
\begin{equation}\label{App2}
\begin{split}
&d_x=\alpha k^3\sin(3\phi)=\alpha(3k_x^2k_y-k_y^3),\\
&d_y=-\alpha k^3\cos(3\phi)=\alpha(3k_y^2k_x-k_x^3).
\end{split}
\end{equation}
Substituting Eq. (\ref{App2}) into Eq. (\ref{Gammaz}), and noting that $d=\alpha k^3$, then
\begin{equation}\label{App3}
\Gamma^z_y=\frac{1}{d^2}\left(d_x\frac{\partial d_y}{\partial k_y}-d_y\frac{\partial d_x}{\partial k_y}\right)=\frac{3\cos\phi}{k}.
\end{equation}
The non-Larmor SHC is given by
\begin{equation}\label{App4}
\sigma^N_{xy}=\sum_{n\mk}f_{n\mk}\frac{3\hbar}{2}\left(+\frac{nq\Omega_0}{\hbar}\frac{1}{\left(\frac{\hbar}{\tau}\right)^2+\Omega_0^2}\Gamma_y^z\right)\frac{\partial\epsilon_k}{\hbar\partial k_x}e^{-t/\tau}.
\end{equation}
Substituting Eq. (\ref{App3}) into Eq. (\ref{App4}) and noting that $\Omega_0=2\alpha k^3/\hbar$, we obtain
\begin{equation}\label{App5}
\sigma^N_{xy}=\frac{9q}{8\pi}\frac{\hbar^2}{m}\int_{k_F^-}^{k_F^+}dk\frac{2\alpha k^4}{\left(\frac{\hbar}{\tau}\right)^2+\left(2\alpha k^3\right)^2}e^{-t/\tau}.
\end{equation}
For the Larmor SHC, $e^{-t/\tau}$ is replaced by $\cos(\Omega_0t)$ and an overall negative sign is added, that is,
\begin{equation}\label{App6}
\sigma^L_{xy}=-\frac{9q}{8\pi}\frac{\hbar^2}{m}\int_{k_F^-}^{k_F^+}dk\frac{2\alpha k^4}{\left(\frac{\hbar}{\tau}\right)^2+\left(2\alpha k^3\right)^2}\cos\left(\frac{2\alpha k^3}{\hbar}t\right).
\end{equation}

 %We note that based on the sign convention in the eigen energy [see Eq. (\ref{EigenE})], the Fermi momentum satisfies $k_F^+>k_F^-$ for %$\alpha>0$. If $(\hbar^2/2m\alpha)\gg\sqrt{4\pi n_F}$, we have $(1/k_F^--1/k_F^+)\approx2m\alpha/\hbar^2$, where %$n_F=(1/4\pi)[(k_F^+)^2+(k_F^-)^2]$ is the hole density~\cite{Loss2005}. We note that the numerical value~\cite{detail} from experimental results yields $(\hbar^2/2m\alpha)\approx1.18(nm^{-1})$ and $\sqrt{4\pi n_F}\approx0.46(nm^{-1})$.


\begin{thebibliography}{99}
\bibitem{tauavg} In this study we focus on the bulk SHC and the boundary condition of a finite sample is not taken into account.  On the contrary, the observed spin-Hall current is due to the spin current near edges of sample. Furthermore, various k points have different Larmor frequencies, which should be considered in the time-averaged SHC. Therefore, the time average accounting for the finite size effect, i.e., $(1/T)\int_0^{T}f(t)dt$ is not used in this paper.
\end{thebibliography}
\end{document}